\documentstyle[12pt]{article}
\title{\bf Exploring Star Clusters Using Str\"omgren $uvby$ 
        Photometry\footnote{Based on data collected at the 
        European Southern Observatory (La Silla, Chile) and 
        the Nordic Optical Telescope, Spain} 
       }
\author{F.~Grundahl $^1$, D.~A.~VandenBerg $^1$, P.~B.~Stetson $^2$\\ 
        M.~I.~Andersen $^3$ and M.~Briley $^4$
\vspace{1cm}\\
{\normalsize $^1$University of Victoria, Department of Physics \& Astronomy,
                 Canada}\\
\normalsize $^2$Herzberg Institute of Astrophysics, National Research 
                Council, Canada\\
\normalsize $^3$Nordic Optical Telescope, Apartado 474, E-38700 Santa Cruz
de la Palma, Spain\\
\normalsize $^4$Department of Physics \& Astronomy, University of Wisconsin, 
 Oshkosh, USA }
\date{\mbox{}}
\begin{document}
\maketitle
\pagestyle{empty}
%
%
\def\bull{\vrule height .9ex width .8ex depth -.1ex}
\makeatletter
\def\ps@plain{\let\@mkboth\gobbletwo
\def\@oddhead{}\def\@oddfoot{\hfil\tiny\bull\quad
``The Galactic Halo: from Globular Clusters to Field Stars'';
35$^{\mbox{\rm rd}}$ Li\`ege\ Int.\ Astroph.\ Coll., 1999\quad\bull}%
\def\@evenhead{}\let\@evenfoot\@oddfoot}
\makeatother
%
%
\def\beginrefer{\section*{References}%
\begin{quotation}\mbox{}\par}
\def\refer#1\par{{\setlength{\parindent}{-\leftmargin}\indent#1\par}}
\def\endrefer{\end{quotation}}
%
%
{\noindent\small{\bf Abstract:} We illustrate the application of high 
precision Str\"omgren $uvby$ photometry to open and globular clusters 
(GCs). It is shown how such data can be used to determine cluster ages, 
independent of distance and only weakly dependent on reddening. We also 
illustrate in detail how $c_1$ index variations point to significant 
star-to-star abundance variations of C and N (through variations in the 
strength of the 3360{\AA} NH band) on the red giant branches (RGBs) of 
all metal-poor GCs observed to date as well as on the main sequences 
of M71 (NGC~6838) and 47~Tuc (NGC~104).
 }
%
%
\section{Introduction}

Str\"omgren photometry offers distinct advantages over broad band 
photometry in the study of field stars and star clusters.  In 
particular the most important features are that it can provide 
estimates of effective temperature, heavy element abundance and 
surface gravity for individual stars. Our present contribution 
will focus on the new insights provided by including $u$ band 
(centered at 3500{\AA}) data for turnoff and RGB stars in open 
and globular clusters.  For a discussion of $u$ band photometry
of horizontal branch (HB) stars in GCs see Grundahl et al. (1999). 
We shall illustrate that, for the study of globular and open 
clusters, the $c_1$ index offers the possibility of determining 
ages independent of distance and nearly independent of reddening.  
Furthermore, there is strong evidence that the variations in the 
$c_1$ index (defined as: $c_1 = (u-v) - (v-b)$; reddening corrected: 
$c_0 = c_1 - $0.2$\,$E$(b-y)$) observed in RGB stars in M13 
(Grundahl, VandenBerg \& Andersen 1998) are due to star-to-star 
abundance variations in N and probably also C.

\section{Observations and Data reduction}

All observations for this project were obtained at the 2.56m Nordic
Optical Telescope and the Danish 1.54m telescope on La Silla using 
the $uvby$ filter sets available there. In most clusters we obtained
samples of 10000--30000 stars. Calibrating stars from the lists of 
Schuster \& Nissen (1988) and Olsen (1983, 1984) were observed on 
several nights during our observing runs. The photometry for the 
clusters was done using the suite of programs (DAOPHOT/ALLSTAR/
ALLFRAME/DAOGROW) developed by PBS (see Stetson 1987, 1991, 1994 
for details).

\section{Illustration of the age dependence of $c_1$}

As part of our program we have observed several old open clusters, 
in order to study their color-magnitude diagrams (CMDs) and eventually
to use them as photometric standard fields. Two of these clusters, 
NGC~2506 and NGC~2243 are known to have a metallicity significantly 
lower than solar, approaching that of 47~Tuc. From previous studies 
of their CMDs it is well known that they span a large range in age, 
and thus we can illustrate how the cluster loci change in a $(v-y, 
c_1)$ diagram (all three clusters have nearly the same reddening, and 
thus we make no corrections for reddening). In Fig. 1a-1c we show 
our calibrated $(v-y, V)$ CMD's for NGC~2506, NGC~2243 and 47~Tuc. 
Note how prominent the sequence of possible equal mass binaries is 
in both NGC~2506 and NGC~2243. (The data for these three clusters 
were obtained during the same observing run, thus differences due 
to variations in filter/detector combinations are ruled out). 
Figure 1d clearly shows how the value of $c_1$ at the cluster turnoff
(at constant metallicity) increases with increasing cluster age. 
There are some obvious advantages to using $uvby$ photometry for 
estimating ages of clusters in this way: the method 1) is independent 
of the cluster distance; 2) does not rely on HB morphology  or number 
of HB stars; 3) is nearly independent of reddening. Assuming ages 
of approximately 2.5, 5 and 12~Gyr for NGC~2506, NGC~2243 and 
47~Tuc, we see that the rate of change in $c_1$ decreases with 
increasing age -- for fairly metal-rich populations this method
is best suited for ages smaller than $\sim$10~Gyr. At lower
metallicities the sensitivity to age of $c_1$ increases. For an 
application of this method to determine the absolute ages of M92 and 
M13, please refer to Grundahl et al. (1999, in preparation)

\section{Abundance variations in {\bf all} globular clusters?}

It has been known for many years that several GCs exhibit 
star--to--star variations in their elemental abundances, 
for both RGB and main sequence stars. In particular, NGC~6752, 
47~Tuc (Cannon et al. 1998) and recently M71 (Cohen 1999a) 
have been studied in detail. These clusters show evidence for 
a bimodality in their CN abundance on the RGB and main sequence.  
Spurred by our finding (Grundahl et al. 1998) that M13 shows 
large star--to--star variations in the $c_1$ index for lower RGB 
stars, we examined our photometry for other GCs, and surprisingly 
found that {\em all} of the observed clusters with [Fe/H]$<-1.1$ 
exhibit the $c_1$ variations (Fig. 2).  For the more metal-rich 
GCs in our sample, M71 and 47~Tuc (Fig. 3), we do not find a large
spread among the giants (probably just due to small sample
size) but very significant variations on the upper two magnitudes
of the main sequence; for fainter stars we cannot distinguish between
photometric scatter and real star--to--star variations. In M71 (not 
shown here) the definition of the total range in $c_1$ is slightly 
problematic due to field star contamination -- there is, however, 
no question that most of the observed scatter is real.  

Figure 2 showns how $c_1$ varies with the luminosity of RGB stars 
in our observed clusters. For each cluster we have marked the 
location of the RGB bump as measured from the RGB luminosity
function (excluding AGB stars). In NGC~288 the ``bump'' is not 
detected, and we have simply estimated its position from comparison
with NGC~362 and NGC~1851 which have a similar metallicity.
For all clusters (except M71 and 47~Tuc, to be discussed below) 
the scatter in $c_1$ remains constant as a function of luminosity at 
fixed metallicity, and as the luminosity increases above the
RGB bump the width of the $c_1$ scatter decreases. 

The most obvious question to ask is what is causing these observed 
$c_1$ variations?. In this respect we first notice that the effect 
exists for all the filter/detector combinations we have used. Second, 
all the RGB stars are so bright (and in most cases have many measurements
in each filter) that the estimated photometric errors for most stars
are reliable and less than 0.01 magnitudes.  Thus the observed scatter 
on the lower RGB is of order 10$\sigma$. We therefore consider the 
scatter to be real and highly significant. It is also evident from 
our data that the scatter among the RGB stars is only significant 
when color indices using the $u$ band are constructed. Thus the most 
obvious candidate for causing these variations is the ultraviolet  
NH band at 3360{\AA}. If this is indeed the case we would expect 
that there should be a correlation between the measured CN strength 
and $c_1$ scatter. 


\begin{figure}
\vspace{11 cm}
\includegraphics{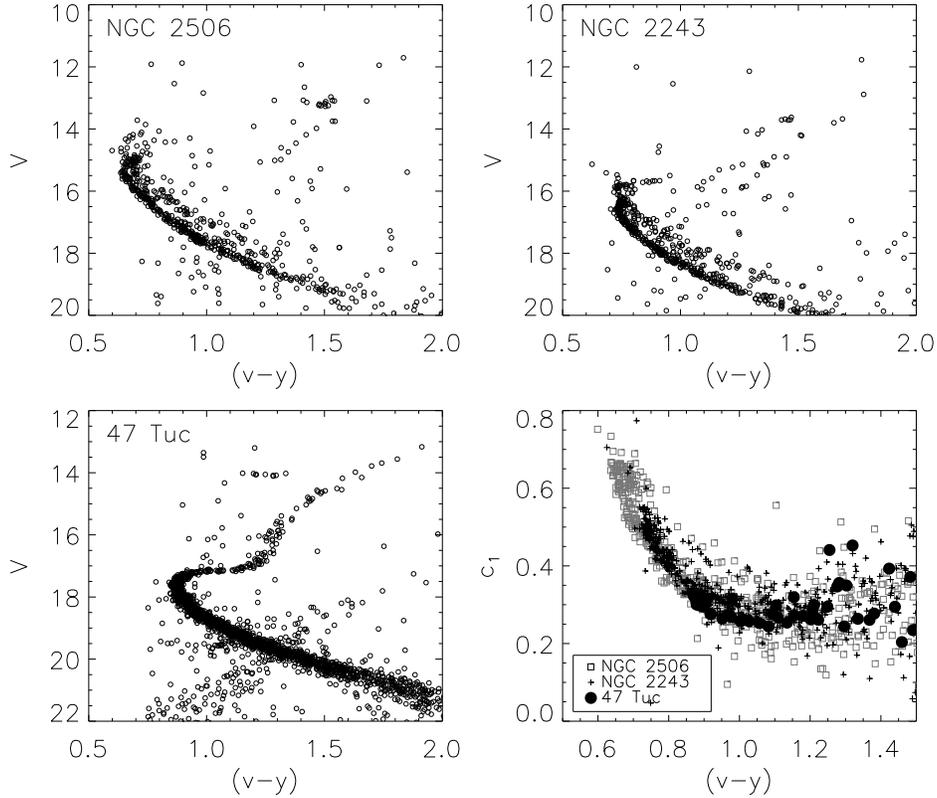}
\caption{Illustration of the age dependence of $c_1$. Panels (a) to (c) 
show the calibrated CMDs for the open clusters NGC~2506 and  NGC~2243 and 
the globular cluster 47~Tuc. Note the ``binary sequence'' in the CMDs of
the open clusters. Panel (d) gives the $(v-y, c_1)$ diagram for the 
three clusters. Due to its higher age the TO point for 47~Tuc (filled 
black circles) appears at lower $c_1$ values. It is also seen that the 
``binary sequence'' disappears in this diagram. (There is a notable 
contamination of the 47~Tuc CMD from stars in the Small Magellanic Cloud).
}
\end{figure}


\newpage


\begin{figure}
\vspace{11 cm}
\includegraphics{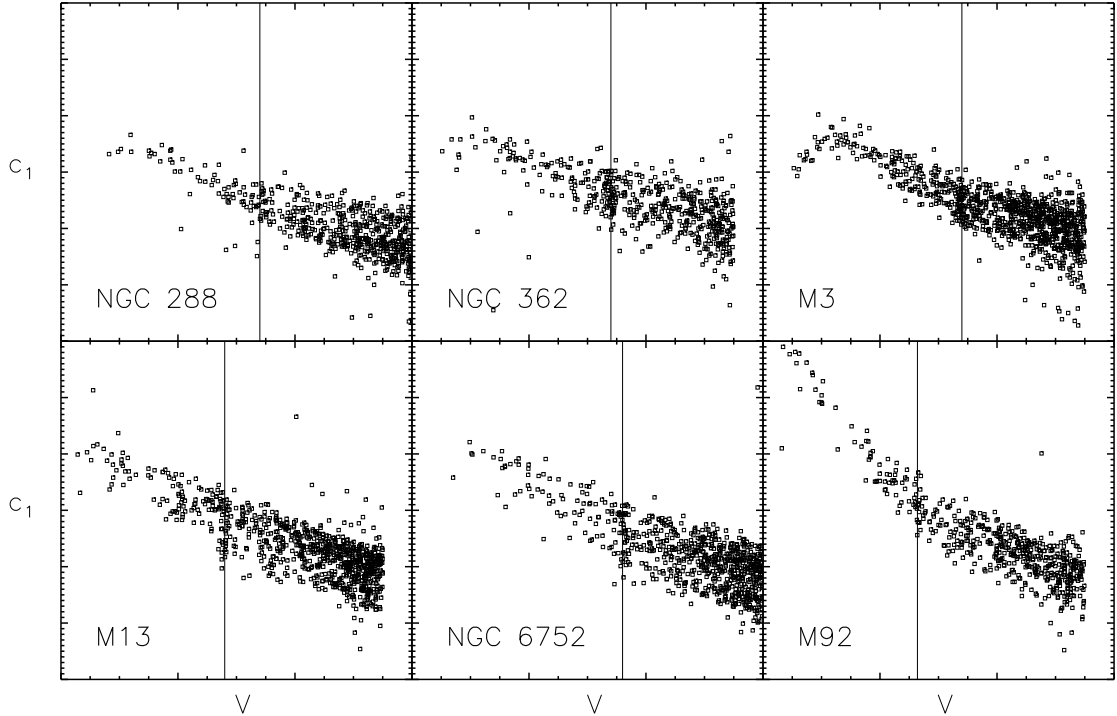}
\caption{A plot showing the scatter in $c_1$ as a function of luminosity
          for the sample of our best observed clusters. Our full
          sample contains 18 GCs. The vertical line
          marks the estimated location of the RGB bump. AGB stars have 
          been excluded from the plot. We see that the scatter decreases
          for luminosities higher than the bump. The scale is the same for 
          all panels}
\end{figure}


Before turning to further empirical 
evidence that variations in the NH strength are causing the $c_1$ 
variations we have calculated Str\"omgren indices for a ``CN strong''
and a ``CN weak'' star and overplotted these on our observations of
 M~13 and 47~Tuc (Fig. 3).  Values for the gravity and temperature 
were taken from a 16~Gyr O enhanced isochrone, similar to that employed 
in the study of 47~Tuc by Hesser et al. (1987). As can be seen the 
models reproduce well the observed width, although there is some 
indication that the actual spread in 47~Tuc may be larger than 
assumed in the models.  The synthetic spectra for RGB and MS stars 
indicate that the cause for the spread in $c_1$ is indeed the 
variations in NH strength. 

To further test this hypothesis we have attempted to match GC stars 
with spectrocsopically measured CN or NH strength with our photometry. 
However, since most of our fields are located close to the cluster 
centers we only have a small amount of overlap. In M92 it is known 
from the study of Carbon et al. (1982) that there are significant 
(factor 10) variations in the abundance of N among faint RGB stars. 
We only have 3 stars with measured NH strength in common with their 
study, of which one is an AGB stars and the other is a bright giant 
in the region of very little (if any) $c_1$ scatter.  The last star, 
however, is classified by Carbon et al. as NH strong, and does indeed 
lie at the upper envelope (as expected from the models) of our $c_1$ 
distribution. In the case of 47~Tuc, we have approximately 38 stars 
in common with the study of Cannon et al. (1998), of which they classify 
19 as CN strong and 19 as CN weak. In Fig. 3, we have overplotted these 
38 stars and there is clearly a separation in $c_1$ of the two groups 
as expected if the $c_1$ scatter is related to N (and probably also 
C) variations.  


\begin{figure}
\vspace{9 cm}
\includegraphics{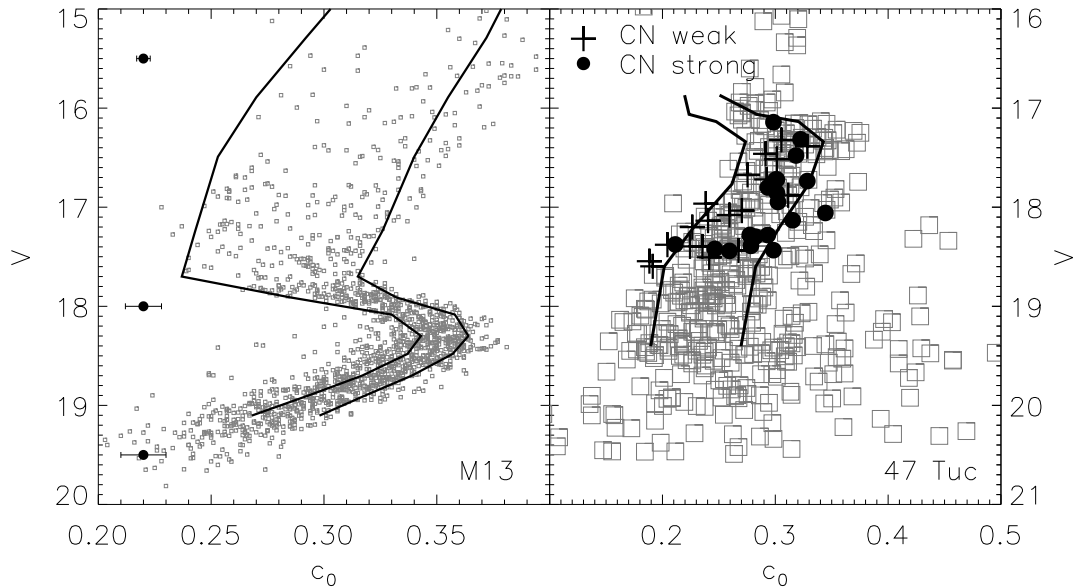}
\caption{A plot showing the scatter in $c_1$ for M13 (left) and 47~Tuc 
         (right) with 
         models overplotted. At fixed luminosity the CN rich stars
         have the highest $c_1$ values. In the right panel the CN 
         weak ($+$) and CN strong ($\bullet$) stars from Cannon et al.
         (1998) are overplotted. }
\end{figure}

In the case of M13, for
which we have the highest quality photometry, there is some evidence
that the scatter is present at $V=18$, only 0.5 magnitudes brighter
than the cluster turnoff. Cohen (1999b) did not report any 
star--to--star variation in the CN strength among turnoff stars in 
M13. We suspect that this is because these stars are so hot that 
most of their CN molecules are destroyed. If this is true, 
observations of NH are needed for the detection of abundance 
variations in these fairly hot stars.  We therefore suggest that 
{\em all} globular clusters have variations in N and probably 
also C in stars as faint as the base of the RGB.  Whether these 
variations are ``primordial'', due to cluster self--enrichment or 
some as yet unknown mixing process ocurring in fairly unevolved 
giants or subgiants remains to be seen. 

\section{Conclusions}

We have demonstrated how the use of Str\"omgren photometry allows the
determination of distance independent ages for both open and globular
clusters. Especially for clusters less than $\sim\,10\,$Gyr the
variation in $c_1$ with age is fairly large, making it relatively easy 
to determine for old open clusters.  It has also been argued that most, 
if not all, globular clusters exhibit evidence for star--to--star 
variations in at least N (and probably C) -- the cause for these 
abundance variations is not yet clear. At luminosities higher than the 
RGB bump we find evidence that the scatter in $c_1$ decreases, 
possibly due to the onset of deep mixing which will dredge up N.

%
%
\section*{Acknowledgements}
Russel Cannon is thanked for providing a machine readable version of 
their spectroscopic results for 47~Tuc.
%
%
 
\beginrefer

\refer Cannon, R.D., Croke, B.F.W., Bell, R.A., Hesser, J.E., 
       Stathakis, R.A., 1998, MNRAS, 298, 601

\refer Carbon, D.F., Langer, G.E., Butler, D., Kraft, R.P., Suntzeff, N.B.,
       Kemper, E., Trefzger, C.F., Romanishin, W., 1982, ApJS, 49, 207 

\refer Cohen, J., 1999a, AJ, 117, 2434

\refer Cohen, J., 1999b, AJ, 117, 2428

\refer Grundahl, F., Catelan, M., Landsman, W., Stetson, P.B. Stetson, 
       Andersen, M.I., 1999, ApJ, Oct. 10 issue, In press.

\refer Grundahl, F., VandenBerg, D.A., Andersen, M.I., 1998, ApJL, 500, 179

\refer Hesser, J.E., Harris, W.E., VandenBerg, D.A., Allwright, J.W.B., 
       Schott, P., Stetson, P.B. 1987, PASP, 99, 739

\refer Olsen, E.H., 1983, A\&AS, 54, 55

\refer Olsen, E.H., 1984, A\&AS, 57, 443

\refer Schuster, W., Nissen, P.E., 1988, A\&AS, 73, 225

\refer Stetson, P.B., 1987, PASP, 99, 191

\refer Stetson, P.B., 1990, PASP, 102, 932

\refer Stetson, P.B., 1994, PASP, 106, 250

\endrefer           
\end{document}